\documentstyle[12pt]{article}
\begin{document}
\begin{center}
\baselineskip 0.70cm
\vskip 0.01in
{\Large \bf New magnetic field induced macroscopic quantum phenomenon 
in a superconductor with gap nodes}
\vskip 0.2in
{\large 
G. Varelogiannis$^{\bf 1,2,*}$}
and {\large M. H\'eritier$^{\bf 3}$} 
\vskip 0.4cm

{\it $^{\bf{1}}$ Department of Physics, National Technical University
of Athens,\\
GR-15780 Athens, Greece}
\vskip 0.05in

{\it $^{\bf 2}$ Max-Planck-Institut f\"ur Physik Komplexer Systeme,
N\"othnitzer Str. 38,\\
01187 Dresden, Germany}
\vskip 0.05in

{\it $^{\bf 3}$ Laboratoire de Physique des Solides, CNRS et 
Universit\'e de Paris-Sud,\\
 91405 Orsay Cedex, France}

\vskip 0.15in

\vskip 0.05in

\vskip 0.1in
\begin{abstract}
\baselineskip 0.80cm

High-$T_c$ superconductivity is unconventional 
because the gap is not isotropic as in simple metals but
has $d_{x^2-y^2}$ symmetry with lines of nodes.
In a fascinating thermal transport experiment 
on a high-$T_c$ superconductor,        
Krishana et al \cite{Krishana}
have reported
mysterious magnetic field induced first order transitions 
from a superconducting state with gap nodes
to a state without gap nodes. We show here that this
is an experimental manifestation
of a novel {\it macroscopic} quantum phenomenon
induced by the magnetic field, qualitatively different from the usual
quantum Hall effects. It corresponds to the {\it quantization of
the superfluid density} in a superconductor with gap nodes due to the
generation of Confined Field Induced Density Waves (CFIDW) in the
node regions of the Fermi surface. 
The Landau numbers $L$ are not sufficient to index
these macroscopic quantum states and the addition 
of a new quantum number $\zeta$ is necessary.
Distinct qualitative implications of
this non-integer $|L,\zeta>$ quantization  
are also evident in a number of recent 
unexplained experimental reports in the cuprates.

\end{abstract}
\vskip 0.0cm
\vskip 0.0cm
\end{center}
$^{\bf{*}}$ e-mail: varelogi@central.ntua.gr
\newpage
\baselineskip 0.90cm
In conventional superconductors the gap is
rather isotropic having s-wave symmetry.
A breakthrough in the study of
high-$T_c$ oxide
superconductors (HTSC) was the demonstration
that instead, the gap is very anisotropic 
of $d$-wave symmetry
taking a zero value in the so called node directions
\cite{vanHarlingen,Tsuei}.
Many of the peculiar properties of HTSC are now simply understood
as resulting from this distinct momentum structure of the gap.
However, understanding the
influence of a {\it perpendicular} to the $Cu-O$ planes magnetic field 
on this ground state is actually a fascinating challenge 
for theorists \cite{Lee1}.
In a remarkable thermal transport experiment,
Krishana et al. \cite{Krishana}                       
have reported that,
when              
a magnetic field is applied
perpendicular to the $CuO_2$ planes,
the thermal conductivity shows sharp first order transitions from 
a field dependent regime characteristic 
of gap nodes \cite{SimonLee} to a field independent regime indicating
the elimination of the nodes.                

The results of \cite{Krishana} stimulated   
a controversial experimental and theoretical debate
which was in fact never closed.
Thermal transport measurements by other groups
\cite{Behnia,Taldenkov,Ando}
have    
confirmed the surprising findings
of  Krishana et al.\cite{Krishana}.
However, they have also reported that the  
phenomenon is not present in all samples \cite{Ando}.
Very high 
quality samples are apparently necessary \cite{Ong2}.
In addition,
strong hysteretic behavior was reported in Ref. \cite{Behnia}
in disagreement with Refs. \cite{Krishana,Taldenkov}.
Our theory provides a simple explanation of this controversy 
and the sample dependence of the occurence of the phenomenon
is a major argument for the relevance of our picture.
On the theoretical side, the phase-transition-point of view
has been immediately adopted by Laughlin
who suggested that this is a transition from $d_{x^2-y^2}$
superconductivity (SC)
to $d_{x^2-y^2}+id_{xy}$ SC \cite{Laughlin} which is nodeless.
Models based on the vortex dynamics have also been proposed 
\cite{Franz} explaining the saturation of the 
field effect but could not account for the ``kink'' and the 
sharp character of
the transition.
 
Indications of a field induced transition to a nodeless
state are also present in other experiments.    
It has been suggested that the
field induced spliting
of the zero bias conductance peak in tunnel
measurements may be the result of such a transition \cite{Deutscher}.
More spectacular are recent measurements by
Sonier et al. \cite{Sonier} of 
the penetration
depth in the presence of a stronger
than usually magnetic field.
They also confirm the elimination of the nodes by the magnetic
field adding a fundamental new element:
{\it The elimination of the nodes 
is accompanied by a substantial
reduction of the superfluid density} \cite{Sonier}. 
Such a reduction cannot be explained by 
a phase transition to a new SC state        
like 
$d_{x^2-y^2}+id_{xy}$.

Moreover, a number of seemingly unrelated experimental puzzles emerged
recently. Neutron scattering experiments reported
the generation of AFM moments in the SC state by a magnetic field applied
perpendicular to the planes \cite{Lake}. The moments appear suddenly
above a critical field and below a critical temperature \cite{Aeppli}.
NMR measurements have confirmed the phenomenon \cite{Mitrovic}.
Scanning tunneling microscopy results report in the presence of 
a perpendicular magnetic field a checkerboard structure
that covers a region around each vortex \cite{Hoffman}. Finally,
measuring both heat capacity and magnetization on an extra
clean sample of YBa$_2$Cu$_3$O$_7$, Bouquet et al. \cite{Bouquet}
not only confirm the presence of a first order transition from a vortex
lattice to a so called vortex glass state, 
but at higher fields they 
observe
a surprising transition from the vortex glass state to a new vortex glass
state which has not found any theoretical explanation so far.

In this Letter we explore an original physical picture which 
is shown to account for {\it all the experiments}.   
We reveal a new magnetic field induced phase transition
from a $d_{x^2-y^2}$ SC state to a 
state in which $d_{x^2-y^2}$ SC {\it coexists with
Confined Field Induced (Spin and Charge) Density Waves (CFIDW) which      
develop in the gap node regions of the Fermi surface (FS)}. 
Field induced density waves have been suggested
\cite{Gorkov,Heritier1, Heritier2,Maki,Yamaji}
in order to explain second order metal-insulator transitions
in $(TMTSF)_2 X$ ($X=PF_6$, $ClO_4$) quasi one-dimensional
synthetic compounds under pressure \cite{TMTSF1}.
The transitions we discuss here are novel and the qualitative
physics of our CFIDW states is novel as well.
We consider an hybrid approach.
Our HTSC system is built of two subsystems:
subsystem I is the Fermi Surface (FS) region covered by 
the superconducting gap and subsystem II is a virtual
normal quasiparticle region created by the magnetic field
and centered in the node
points of the FS.
The 
CFIDW will eventually develop in region II because of the 
orbital effect of the field in this region. 
The orbital effect of the field in region I
induces vortices that are assumed to be irrelevant.
The Zeeman effect
is not considered here for simplicity, however
in region II it can be shown
to further stabilize our CFIDW states while in region I it is
negligible.
The relative momentum extension of regions I and II  
is unknown and  
fixed by the energetic competition of CFIDW with SC.
The above picture has some similarity with
the partially depaired 
Fulde-Ferrell-Larkin-Ovchinikov state
\cite{FF,LO} where the normal state is created by the magnetic field
over a portion of the Fermi surface and competes with 
superconductivity.

In HTSC the gap is $d_{x^2-y^2}$ with nodes in the 
$(\pm\pi,\pm\pi)$ directions where region II is centered.
In region II {\it we necessarily have open FS sheets}.
Therefore, we can             
write the dispersion of subsystem II  
in the form:        
$
\xi^{II}_{\bf k}=\upsilon_F(|k_1|-k_F)-2t_2\cos (k_2/X)-
2t'_2\cos (2k_2/X)
$
where $X$ is the unknown momentum extension of region II
(see Fig. 1a).
$ k_1$ is along the
$(\pm\pi,\pm\pi)$ directions perpendicular to the open FS sheets of
subsystem II,
and $k_2$ is perpendicular to $k_1$ and therefore along
the open FS sheets where we keep only two harmonics
without influence on the generality of the results.

We are able to calculate
explicitely the spin and charge susceptibility of subsystem II
in the presence of the magnetic field                  
exploiting methods developed for the
study of $(TMTSF)_2 X$ compounds 
\cite{Heritier1,Heritier2}.
The details of the 
calculation and a             
discussion
with various                           
electronic dispersions in II
will be given elsewhere. We report here the simplest
results sufficient to illustrate 
the surprising physics of the system.
Constraining the nesting in the $(\pm\pi,\pm\pi)$ directions
one can show that
a first order       
field induced density wave gap in region II 
is given by 
$
\Delta_{DW}=W\exp \bigl\{ - [g N(E_F)I^2_L(X)]^{-1}\bigr\}
$
where
$$
I_L(X)=\sum_{n}J_{L-2n}\biggl({4t_2 X\over eH\upsilon_F}\biggr)
J_n\biggl({2t'_2 X\over eH\upsilon_F}\biggr)
\eqno(1)
$$
Here $J_n(x)$ are Bessel functions, $L$ is the index of 
the Landau level configuration,
$e$ is the charge of the electron, $H$ the magnetic field, 
$N(E_F)$ the density of states at the Fermi level (in region II), $g$
a scattering amplitude of Coulombic or phononic origin, 
$W$ the bandwidth in the $(\pi,\pi)$ direction
and $\upsilon_F$ the 
Fermi velocity. 
Higher order gaps are not reported here for sake of clarity.

CFIDW states will develop only if 
the absolute free energy gain of the system
due to the opening of a CFIDW gap in 
region II is bigger than that
lost by the elimination of SC  
from this region. 
This condition implies the following inequality:
$$
I^2_L\biggl(k_F\sin(Z\pi/2)\biggr) >  
{2\over g N(E_F)}\biggl[\ln{2W^2\pi Z\over \Delta^2_{sc}
[\pi Z - \sin(\pi Z)]}\biggr]^{-1}
\eqno(2)
$$
Z is a dimensionless number varying from $0$ to $1$ that represents
the relative extension of the CFIDW states over the FS             
and is related to $X$ by $X\approx k_F \sin (Z\pi/2)$.
$\Delta_{sc}$ is the maximum value of the $d_{x^2-y^2}$ SC gap.
The CFIDW state must be {\it confined in momentum space}
with a DW 
gap smaller or equal to the absolute superconducting gap in the 
borders of region II. We therefore have:
$$
I^2_L\biggl(k_F\sin(Z\pi/2)\biggr)\leq
\biggl[g N(E_F)\ln {W\over \Delta_{sc}\sin(Z\pi/2)}\biggr]^{-1}
\eqno(3)
$$

The equality in (3) 
fixes                
the relative extension $Z$ of the CFIDW (for each $L$)
and therefore    
$I^2_L\biggl(k_F\sin(Z\pi/2)\biggr)$ which then 
fixes the CFIDW gap $\Delta_{DW}$ 
and the critical temperature $T_{DW}$ at which the CFIDW forms.
A graphic solution for Z is shown in Fig. 1b. 
Surprisingly, there are two possible values of Z for a given Landau level
configuration.
Therefore,
the Landau numbers $L$ are not 
sufficient to index our 
quantum states. A {\it new quantum number} $\zeta$, associated with the 
two possible momentum extensions $Z$ for each $L$ configuration must
be added.
Physically $\zeta$ will index {\it the quantization of the superfluid
density}. In fact, each quantized value of Z corresponds to a different
relative extension of the SC region over the FS
and therefore to a different density of superfluid carriers.

Using for our parameters values extracted
from the experiments on HTSC and assuming a conventional
scattering $gN(E_F)\approx 1$ we obtain
results like those reported in Figures 2 and 3 in remarkable agreement
with the data of
\cite{Krishana} and \cite{Sonier}.
In Fig. 2 is reported the
dependence of $T_{DW}$
on the magnetic field
in 
the various $|L,\zeta >$ configurations.
We also
plot in this figure the 
experimental
points of refs \cite{Krishana} and \cite{Sonier}.
In Fig. 3 we plot the corresponding
magnetic field dependence of the accessible relative
extensions $Z$ of the CFIDW in each $|L,\zeta>$ configuration
for the same parameters.
A quantitative fit of the experiments as in figure 2
establishes that
the orders of magnitude of the involved parameters
are compatible with the experimental data.          
More importantly,
there are distinct qualitative 
experimental facts which 
strongly support our picture. As one can see in
Fig. 2, the $T_{DW}$ versus critical magnetic field profile
of the data in \cite{Krishana} (circles)
show a {\it  ``reptation''} shape.
The first two points have a bigger field slope than the
next two points and so on. 
To the best of our knowledge, no explanation of the
reptation behavior has been reported so far.
Within our analysis this
``reptation'' profile
is due to the {\it quantization} of the CFIDW states.
Each slope corresponds to a different $|L,\zeta>$ quantum configuration
of the system. The higher the field is, the smaller is the field
slope in agreement with the experiment.
Moreover, if one associates the vortex solid to vortex glass
transition with the SC to SC+CFIDW transition, one naturally
understands {\it the unexplained vortex glass I to vortex glass II
transition of Bouquet et al \cite{Bouquet}
as a transition between two different $|L,\zeta >$ configurations
of the SC+CFIDW state}. The higher field $|L,\zeta >$ state
corresponds to a higher relative extension of the
CFIDW state and therefore a less rigid vortex structure as
is experimentally observed.
Alltogether,
this provides a microscopic
viewpoint for the origin of the spin-glass states.

The remarkable sensitivity of
the occurence of this phenomenon
on sample quality \cite{Ando,Ong2} further corroborates our picture.
Within our analysis,
a small magnetic field affects
the large SC gap of HTSC creating CFIDW's
in the node region {\it because of momentum confinement}
which
constrains the 
cyclotron orbits to be large in real space
and consequently the involved flux is large as well.
The occurrence of the CFIDW's {\it requires very            
clean samples} because the mean free path must be bigger
than the cyclotron orbits.
Samples which may appear of high quality
from the usual criteria (width of the SC transition
or magnitude of the SC $T_c$) may not be sufficiently
clean to support large cyclotron orbits and show the CFIDW states.
This explains the puzzling sample dependence of the 
occurence of the phenomenon.
The minimum field at which the phenomenon is observable
is also limited by sample defects because if fields are too
small, momentum extensions of the CFIDW are also too small
(see figure 3) and consequently the required cyclotron 
orbits are too large.
Our $d_{x^2-y^2}$ to $d_{x^2-y^2}+CFIDW$ transition 
appears only
above 1 Tesla in \cite{Behnia} while it is already present at 
0.6 Tesla in \cite{Krishana} because samples in 
\cite{Krishana} 
are cleaner admiting bigger cyclotron orbits.   
Because all $|L, \zeta >$ configurations are 
nearly degenerate for the total system,
the occurence of magnetic hysteresis on a dynamic probe like
thermal transport will depend on the exact conditions 
of the magnetic cycle and on sample quality as well.
In the magnetic cycle of Ref. \cite{Behnia}
showing hysteresis,
the field orientation is reversed when the field maxima are reached
(the field passes through zero),
while in Refs. \cite{Krishana,Taldenkov} 
this is not the case and hysteresis
is absent. 

Our analysis is also the first to
establish a natural relationship between the thermal
transport \cite{Krishana} and penetration depth \cite{Sonier}
puzzles. In \cite{Sonier} CFIDW develop at a much
higher $T_{DW}$ for a given field than in
\cite{Krishana} because the system is cooled 
in the presence of the magnetic field. By field cooling              
the sample, the first accessible $|L, \zeta >$
configuration is the one with the higher $T_{DW}$ which 
also corresponds to the higher $Z$ (cf. figure 3). 
On the other hand, in the experiment of \cite{Krishana},
the temperature is kept constant when the field varies.
Not only we reproduce simultaneously the experimental $T_{DW}$
versus field profiles of both
\cite{Krishana} and \cite{Sonier},
but we also account for the reduction
of the superfluid density reported in \cite{Sonier}.
As one
can see in figure 3, in the regime $4$ - $6$ T
explored in \cite{Sonier}, the CFIDW occupy
about a quarter of the FS ($Z\approx 0.25$ to $0.35$) which
is in very good agreement with the $\approx 15\%$ to $25\%$
reduction of the superfluid density reported in \cite{Sonier}.
Furthermore,
our CFIDW states correspond to a real
space pattern that has all the characteristics of the one observed
by STM \cite{Hoffman}.

Finally,  
in agreement with all the 
experiments, our CFIDW states appear
{\it only for fields perpendicular
to the planes} because for
fields parallel to the planes cyclotron
effects are irrelevant.

\newpage

\vskip 2cm
\begin{center}
{\Large \bf Acknowledgement}
\end{center}
\vskip 1.5cm

We are grateful to J. Friedel and P. Fulde
for stimulating and encouraging
discussions and a critical reading
of the manuscript.
We also acknowledge
discussions with
Y. Kudasov, M. Lang, P.M. Openneer and P. Thalmeier and
support from
the Max-Planck Gesellschaft.

\newpage

{\Large \bf Figure Captions}
\vskip 1.5cm

{\bf Figure 1:} a):
Schematic view of the SC-CFIDW competition.
b):
Graphic solution of            
(3) in the $L=1$ configuration and a field of 3 Tesla.
Two different relative extensions Z 
($X\approx k_F \sin (Z\pi/2)$) of the CFIDW are 
possible. 
\vskip 0.5cm

{\bf Figure 2:} (color): Critical temperature $T_{DW}$
versus critical field for the formation of a CFIDW state in
the different quantum configurations $|L,\zeta>$:
$L=4$ (orange), $L=3$ (red), $L=2$ (blue),
and $L=1$ (green). In all $L$ configurations, 
full lines and dotted lines
correspond to the two different $\zeta$ configurations (full lines
to the higher $Z$ solution).
The $L=0$ lines are not shown for clarity.  
The open circles are the corresponding
experimental points of \cite{Krishana}
and the open squares the experimental points of \cite{Sonier}.
\vskip 0.5cm

{\bf Figure 3:} (color): Relative extensions $Z$ versus 
magnetic field for various quantum configurations $|L,\zeta>$.
The different colors and line-styles correspond to the same 
$|L,\zeta >$ as in  Fig. 2.

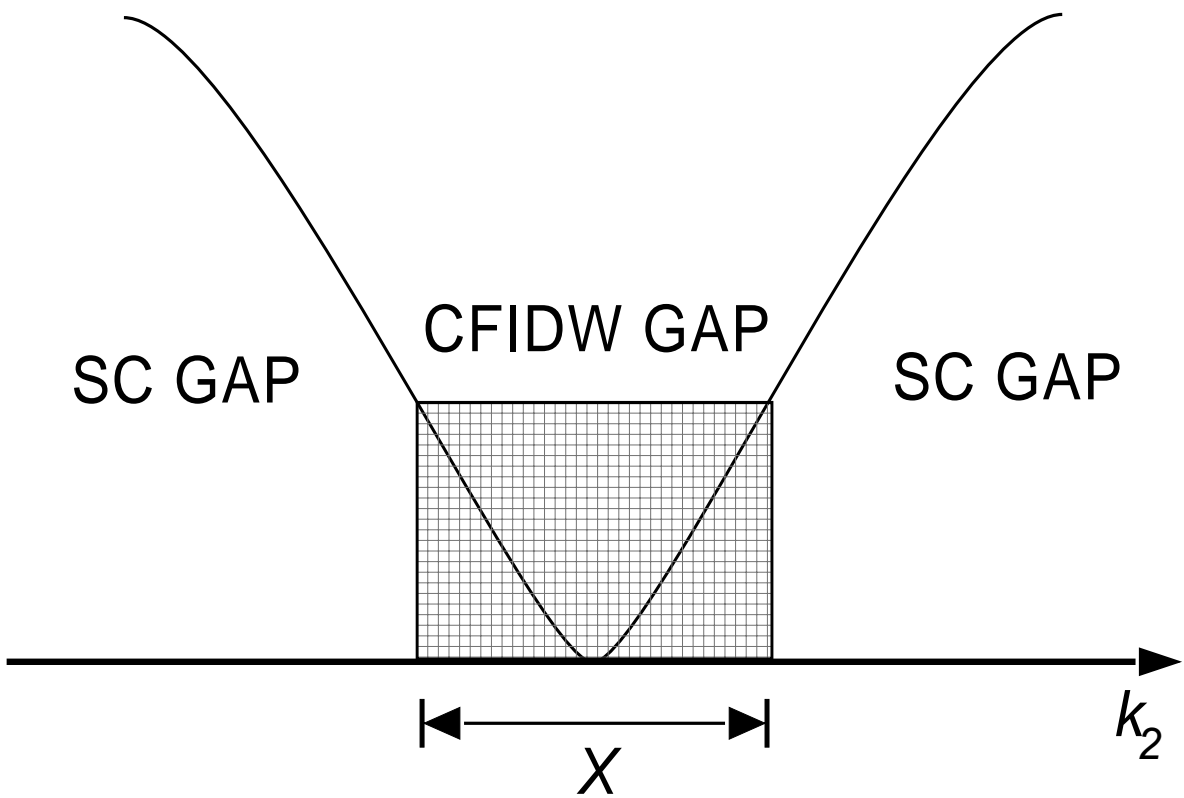
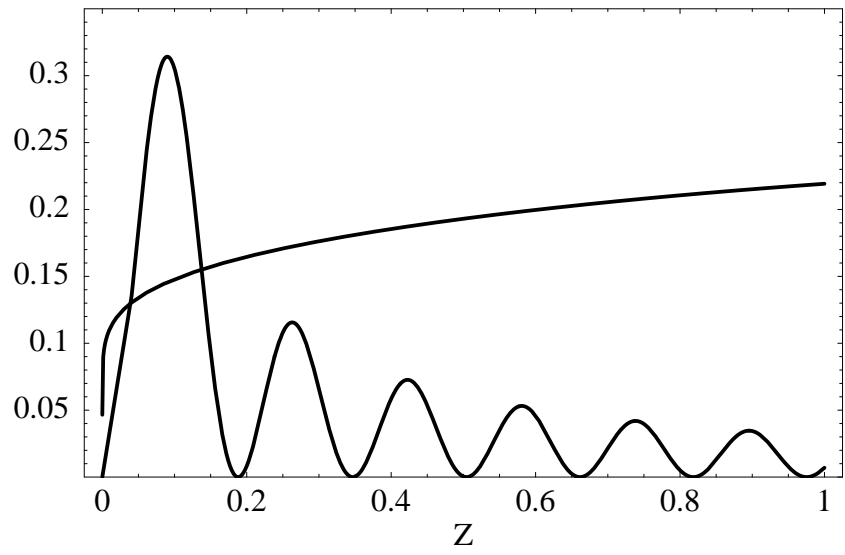
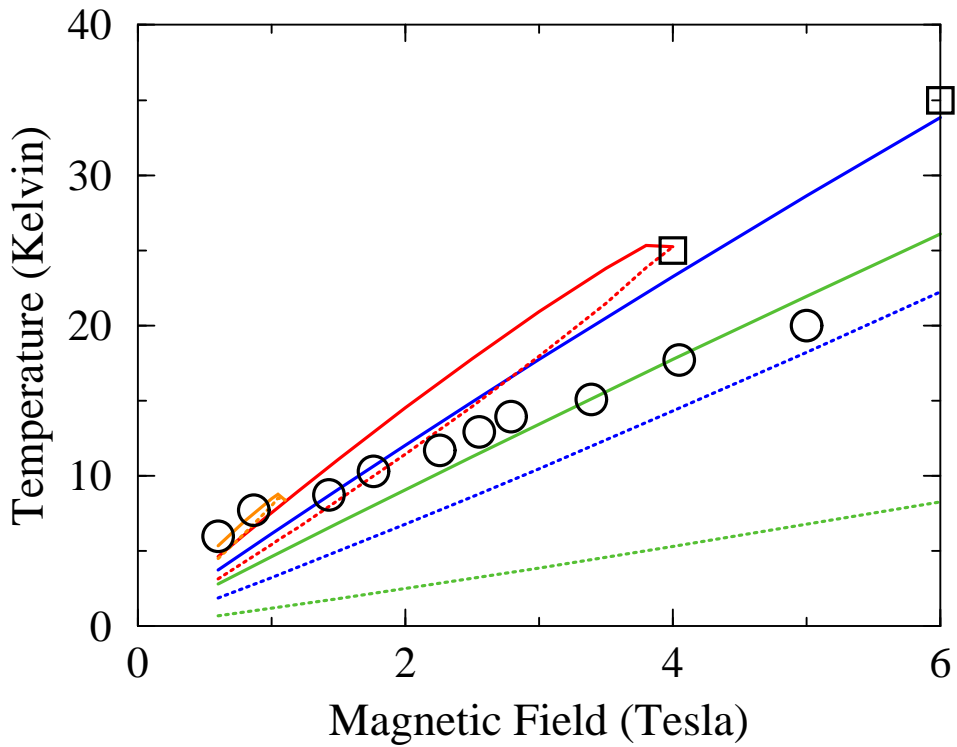
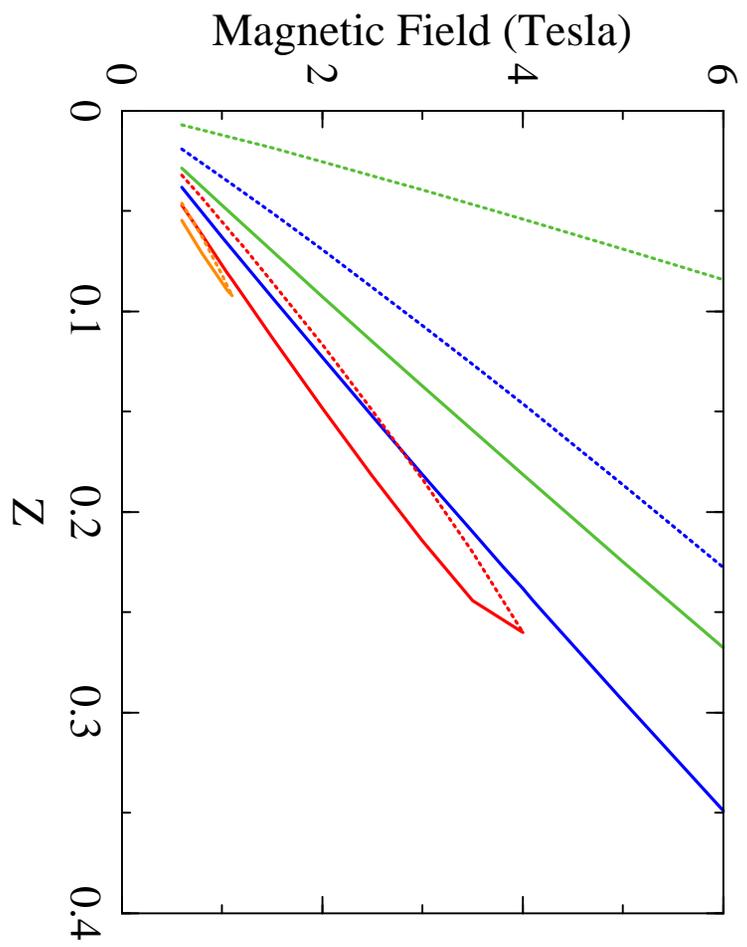
\end{document}